# SOFIA Astronomy and Technology in the 21$^{st}$ Century[1]

Alfred Krabbe and Hans-Peter Röser

*German Aerospace Center, Institute of Space Sensor Technology, Berlin*
http://www.sofia.dlr.de

**Abstract**

SOFIA, the Stratospheric Observatory For Infrared Astronomy mounted onboard a Boeing 747SP will open a new era in MIR/FIR astronomy. Starting in 2002, SOFIA will offer German and American astronomers a unique platform, providing regular access to the entire MIR and FIR wavelength range between 5µm and 300µm part of which is otherwise inaccessible from the ground. SOFIA's 2.7m mirror and optimized telescope system combines the highest available spatial resolution with excellent sensitivity. SOFIA will operate in both celestial hemispheres for the next two decades. In this paper we present an overview of the SOFIA project and the science that this observatory will be able to address.

## 1. Introduction

The infrared (IR) spectral band (NIR: 0.9–3µm, MIR 3–30µm, FIR: 30–300µm, Submm: 300µm–1mm) spans a range about 10 times wider than the visible. During the past 30 years the scientific development of the infrared band was dominated by technological developments of incoherent photon sensitive detectors and coherent antenna-mixer structures. Today, IR astronomy has evolved into a major field of astronomical and cosmological research, and the ongoing activities demonstrate the amazing scientific potential that the IR band bears. However, despite all the technological effort, less than half of the IR waveband is directly accessible from ground with sufficient observing efficiency. Most of the IR radiation from outside our planet is absorbed in the atmosphere, mostly in layers of water vapor, $CO_2$, ozone and other molecules. Figure 1 shows the atmospheric transmission in the IR band from the ground. The wavelength range between 30µm and 300µm is completely wiped out by the atmosphere. In striving to overcome these absorbing layers, IR observatories have been set up on high mountains (e.g. Mauna Kea, Hawaii, 4200m), on airborne observatories (e.g. Lear Jet and Kuiper Airborne Observatory (KAO), 12 – 15km), on balloons (e.g. THISBE and Golden Dragon, 30km), and on satellites in space (e.g. IRAS, ISO). Observatories on high mountains benefit mostly in the NIR and in part of the MIR. Access to the full IR range only begins above the tropopause at about 12km altitude (see Figure 2) where 85% of the range is accessible.

---

[1] Invited talk to be published in: *Reviews in Modern Astronomy,* Vol.12, R.E. Schielicke (ed.), Hamburg: Astronomische Gesellschaft 1999

Comparison of the astronomical potential of the different platforms (aircrafts, balloons, and satellites) weighted by cost per useful photon, scientific flexibility, and spatial and spectral resolution, favors the aircraft. It can easily cover both hemispheres and trace phenomena like stellar, lunar and other occultations, which are only locally observable. It can quickly react to other transient phenomena like comets or novae and supernovae. The instrumentation flown will always be up-to-date and at the forefront of the available technology. Changing instrumentation will be straightforward and optimization of the instrument performance during the observation is possible. The predecessors of SOFIA, the Lear Jet and the KAO, have proven the concept of an aircraft observatory to be very flexible, long lasting, economic and technologically mature for most research areas in the IR. With SOFIA the dream of a big flying observatory is now becoming true.

Figure 1. Atmospheric absorption of infrared radiation on the ground, scaled from 0 to 100%. Responsible for the absorption are water vapor, $CO_2$, and other molecules in atmospheric layers below about 12km. The upper panel shows background continuum radiation emitted by a warm telescope and the atmosphere.

SOFIA is a technological challenge and an unprecedented opportunity for astronomy. Section 2 summarizes the history of the project. An account on the optical and mechanical layout of the telescope, the aircraft and the home base is given in section 3. The operation phase and the education and public outreach are briefly described in section 4 and 5. Section 6 summarizes the expected performance while section 7 briefly discusses the scientific potential of SOFIA. A brief account on the SOFIA first light instruments is given in section 8.

## 2. History of the SOFIA Project

The U.S. aircraft mounted astronomy program began 1969 with a 30cm telescope onboard a Lear Jet. It was followed by the famous Kuiper Airborne Observatory (KAO), a 92 cm Cassegrain telescope onboard a Lockheed L200 Starlifter (C141). All aircrafts were operated by the *NASA Ames Research Center* south of San Francisco. The KAO had 60 to 80 flights per year, each one about 7 hours long. It was operational for 22 years until 1996. For a review of the KAO see Larson (1994). The KAO opened the MIR and FIR window to the universe. Many discoveries were made with this instrument including the detection of interstellar water in comets and the rings of Uranus. Although the KAO was an all American project, two German groups, at the *MPI für Radioastronomie* in Bonn and at the *MPI für extraterrestrische Physik* in Garching contributed their own instrumentation and were invited to participate. The overwhelming scientific success of the KAO and the very successful American-German collaboration were the most important arguments in the final decision for the bilateral SOFIA project.

Figure 2
Comparison between the atmospheric trans-mission on Mauna Kea, one of the best infrared sites on the ground, and with SOFIA.

**Table 1**     Summary of basic SOFIA characteristics

| | | |
|---|---|---|
| Project | Operated by | USRA for NASA, DLR |
| | Development began | January 1997 |
| | First light | 2002 |
| | Operation | 20 years |
| Telescope | Aperture | 2.5m (2.7m) |
| | Wavelength range | 0.3 - 1600μm |
| | Spatial resolution | 1-3 arcsec for $0.3 < \lambda < 15\mu m$ |
| | | ($\lambda/10$ arcsec) for $\lambda > 15\mu m$ |
| Operation | Number of flights | ca. 160 per year |
| | Operating altitudes | 12 - 14 km, 39000 - 45000 ft |
| | Nominal observing time | 6 - 9 hours |
| | Crew | ca. 12: pilots, operators, technicians, and scientists |
| Instrumentation | Instrument changes | 15 - 25 per year |
| | Research flights | ~150 per year |
| | In-flight access to instrumentation | continuous |

In 1984, on the 10[th] anniversary of the KAO, the idea for a new and bigger aircraft mounted telescope was presented and a Boeing 747 was envisioned as the carrier of a 3m-class telescope. In 1987, the *Red Book* was presented which served as the *Phase A System Concept Description* and has been the basis for the project since ever. At that time it was agreed that the German contribution would be 20% of the total cost, including manufacturing, operation, and personnel. The telescope system at the heart of the observatory would be developed and delivered by Germany. The system aperture was set at 2.5 m, which requires a 2.7 m primary mirror diameter for field-of-view, chopping, and diffraction effects. Despite good progress in the technical studies, political goodwill slowed down. The reunification in Germany and severe budget cuts at NASA threw the project back by more than 5 years. For budget reasons NASA decided to privatize the observatory and contract the development and operation of the observatory out to the *Universities Space Research Association* (USRA). The observatory will be built in compliance with the regulations of the Federal Aeronautic Administration (FAA). In December 1996, NASA and DARA (now DLR) finally signed the memorandum of understanding (MoU) on the development and operation of SOFIA. DLR awarded a contract to MAN and Kayser-Threde to develop the SOFIA telescope, and NASA awarded a contract to the Universities Space Research Association (USRA) to develop the rest of the observatory and to operate SOFIA. The basic characteristics of SOFIA are listed in Table 1.

Figure 3. The SOFIA telescope is designed as a Cassegrain system with two Nasmyth foci, the nominal IR focus and an additional visible light focus for guiding. IR light is reflected off the upper tertiary mirror, which has a dichroic coating. The dichroic can be replaced with a fully reflective mirror if guiding with the focal plane guider is not required.

## 3. SOFIA

### 3.1 Optical System

The optical layout of the SOFIA telescope is shown in Figure 3 and summarized in Table 2. It is basically a Cassegrain system with a parabolic primary and a hyperbolic secondary. The secondary mirror is attached to a chopping mechanism providing chop amplitudes of up to ± 5 arcmin at chop frequencies between 0 and 20 Hz, programmable by either a user supplied analogue or TTL curve or by the telescope control electronics. A flat tertiary mirror reflects the beam into the infrared Nasmyth focus, 300 mm behind the instrument flange. If the tertiary is replaced by a dichroic mirror, the transmitted optical light is reflected by a second tertiary 289.2 mm behind the dichroic and sent to the visible Nasmyth focus. There it is fed into a focal plane guiding camera system (FPI). Independent of the FPI there are two other imaging and guiding cameras available: the wide field imager (WFI) and the fine field imager (FFI). Both of these cameras are attached to the front ring of the telescope. Details of their optical performances are summarized in Table 3.

Figure 4.
The primary mirror at REOSC, Paris, during the light-weighting process on a working bench. Milling into the backside of the blank leaves a honey-comb structure with a wall thickness of 7mm. At completion, 85% of the glass will have been removed from the mirror.

**Table 2** Optical parameters of the SOFIA telescope

| Parameter | Value |
| --- | --- |
| Entrance pupil diameter | 2500 mm |
| Nominal focal length | 49141 mm |
| Unvignetted field of view | ± 4 arcmin for chop amplitudes up to ± 5 arcmin |
| Aperture stop location | Secondary mirror |
| Aperture stop diameter | 352 mm |
| Primary free optical diameter | 2690 |
| "      focal length | 3200 |
| "      conic constant | -1 |
| "      center hole diameter | 420 mm |
| "      material | Zerodur (Schott) |
| "      mass | ~850 kg |
| Secondary radius | 954.13 mm |
| "      conic constant | -1.2980 |
| "      material | Silicon Carbide |
| e (Bahner, 1967) | 2754 mm |
| e+g (Bahner, 1967) | 6849 mm |

### 3.2 Telescope structure

One of the design goals of the SOFIA observatory was to make every component as light as possible in order to maximize the useful observing flight duration. Every metric ton of weight saved translates into more than 8 minutes of additional observing time. Saving weight on the telescope itself is particularly important. All additional weight on e.g., the mirror support structure, counts fourfold: It is balanced by the counterweights on the opposite side of the telescope bearing and the telescope itself is balanced within the aircraft by counterweights in the front of the aircraft. Although the telescope as shown in Figure 5 only weighs about 17 metric tons, its structure still has to be very stiff. Therefore, most of the telescope structure, including the mirror support, the telescope frame, the secondary support, and the Nasmyth tube, is made of carbon fiber. The lowest eigenfrequencies of the telescope bending modes are above 35 Hz.

**Table 3** Optical data of the visual imaging and guiding cameras

| Parameter | Value |
|---|---|
| CCD chip size | 1024 x 1024 pixel |
| Pixel size | 14μm x 14μm |
| FFI optical system | Schmidt-Cassegrain |
|     primary diameter | 254 mm |
|     night sensitivity | 13 mag (after 2.1 sec) |
|     pointing stability | 0.35 arcsec |
| WFI optical system | Petzval |
|     front lens diameter | 70 mm |
|     night sensitivity | 8 mag (after 0.5 sec) |
|     pointing stability | 1.8 arcsec |
| FPI optical system | Eyepiece + commercial CONTAX 1.4/85mm lens |
|     night sensitivity | 16 mag (after 1.7 sec) |
|     pointing stability | 0.035 arcsec (nominal) |

At the time of writing the principal design of the telescope at the industry consortium MAN in Gustavsburg and Kayser-Threde in Munich is nearly complete. The critical design review for the telescope system is scheduled for spring 1999. Long lead items like the primary are already at an advanced stage (Figure 4).

Figure 5. Current design of the SOFIA telescope structure. The primary and secondary mirrors (right), together with the carbon fiber support structure, are located in the cavity and will be exposed to about –50°C during flight. The vertical ring (middle) is structurally integrated into the bulkhead (see Figure 7). The bearing, motors, encoders, in edition to the vibration isolation system are mounted around the Nasmyth tube behind the ring. The Nasmyth tube ends at the science instrument flange (left). The box below the flange contains the focal plane imager; the counter weights are located above the flange. The telescope as shown here weighs about 16 metric tons.

### 3.3 Aircraft

The aircraft for SOFIA is a used Boeing 747 SP, slightly shorter than the normal version. Bought in spring 1997, it is currently being modified at the Raytheon Company in Waco, Texas. Figure 6 shows a picture from a test flight in fall 1997. The location of the telescope cavity in the rear part of the aircraft has been painted black. Initially the location of the telescope was foreseen in the front

Figure 6.
SOFIA aircraft during a series of test flights in fall 1997 prior to modification at the Raytheon company in Waco, Texas. The location of the cavity is painted black.

section of the aircraft. However, it turned out that such a solution would have required a very costly pressurized tunnel to allow people to pass from one side of the cavity to the other, and two pressure bulkheads rather than only the one required for the aft telescope location

Figure 7. Cut-away view into the SOFIA aircraft.

Figure 7 shows a cut-away view through the aircraft. The telescope is located in a cavity in the rear, which is sealed off from the pressurized, warm cabin area in the front by a bulkhead wall. The bulkhead wall supports all the weight of the telescope. The telescope drive system, counterweights, most of the Nasmyth tube and the science instruments are located within the cabin and are accessible during flight. Access to the cavity is provided through a door in the aft section when the aircraft is on the ground. This section also contains the devices for pre-cooling the telescope prior to each observing flight and pressurizing the cavity with dry $N_2$ during warm up after a flight.

Figure 8.
The barrel door seals the telescope cavity from the outside. It is opened when the cruising altitude has been reached. This CAD drawing shows that the opening is kept as small as possible in order to minimize air turbulences within the cavity. Minimizing wind loads on the telescope will ensure a good tracking quality.

The cavity is sealed from the outside by a slightly recessed barrel door, which is shown in Figure 8. The door opens as soon as the cruising altitude has been reached, but keeps the opening as small as possible. Wind tunnel tests have shown that an optimized outer shape of the door can more or less maintain the laminar airflow across the hole. However, some turbulent wind loads on the telescope structure will still remain. These wind loads will be compensated for by the telescope drive system.

For balancing reasons, most of the heavy telescope control electronics, power supplies and heat exchanger are located in the front of the aircraft. The observers and the telescope operation crew are located in the section between the wings. On the first and second floor some space will be reserved for visitors (see below).

### 3.4 Mission operation center

SOFIA's home base will be at the SOFIA Science and Mission Operation Center (SSMOC) at NASA Ames Research Center, Moffett Field, south of San Francisco. It is located in a building with a hangar attached to it, where SOFIA will be maintained and kept between flights. The SSMOC staff will consist of about 80 people including scientists, engineers, administrators, and associated services 20% of the crew will be recruited by the DLR from Germany. The SMOC provides laboratory space for various tasks, such as checking and calibrating the telescope and its components, coating the mirrors, and preparing flight schedules. Additional lab space and a telescope simulator will be available to prepare and check science instruments before they are attached to the telescope. SSMOC will also provide office space for visiting scientists. With everything under one roof we expect that SOFIA's home base will quickly evolve into a lively science center where people from different disciplines work hand in hand. Figure 9 gives an impression of the SSMOC building and hangar as it will look about three years from now.

Figure 9. The SOFIA Science and Mission Operation Center (SSMOC) includes all facilities to run SOFIA: A hangar for the aircraft, several labs for optical and mechanical tests and maintenance, a coating facility, laboratory space for preparing and testing visiting science instruments, offices and conference rooms for scientists, administration, flight planning, data reduction and archiving, as well as other services.

Figure 10.
Example of an optimized flight track with the KAO. This picture represents a 7 hour flight at an altitude of between 12 and 14 km. SOFIA flight tracks will look similar. They may, however, cross the Canadian border more frequently. The tropo-pause comes further down at higher latitudes, allowing lon-ger flights at lower altitudes.

### 4. Operation

An airborne observatory poses several constraints on the observing strategy. The length of one flight is limited to about 7-9 hours, implying that the night is always short. Most observing flights will start and end at the same airport: Moffett Field for the Northern Hemisphere and Christchurch, New Zealand for the Southern Hemisphere. The maximum observing time per object is therefore limited to 3.5 hours per flight. The unvignetted elevation range of the telescope is 20° to 60°, which excludes zenith observations. These constraints require detailed flight planning to maximize of scientific information return per target, or per target list, and to avoid dead legs. Figure 10 shows an example of a flight track from the KAO.

SOFIA will be required to operate for at least 960 successful flight hours per year. Assuming a technical success rate of the observatory of 80% and 7 hours per flight of usable astronomical time, the observatory will be flying every other day. With such a tight schedule instrument change rates are limited on average to about one per 2 weeks. The mirrors will be recoated twice per year. SOFIA will be sent over to Germany once per year for a PR event (e.g. Internationale Luftfahrt Ausstellung, ILA, Berlin) and for maintenance checks.

The German Aerospace center (DLR) will organize the German 20% share of the observing time for the German astronomical community, independently of the US. Astronomers at German research institutions will be able to apply for observing time with all publicly available American or German instruments. The DLR will install a peer review board and will schedule the observing time in close collaboration with the US side. The DLR is planning to set up a SOFIA of-fice, which will serve as the point of contact for the German astronomical community as well as for the German SOFIA team members at NASA Ames.

This office will be the successor of the existing SOFIA project office at the DLR Institute of Space Sensor Technology in Berlin.

## 5. Education and public outreach

The American SOFIA project includes a program for education and public outreach (EPO) to foster the public interest and knowledge in astronomy. SOFIA is exceptional in that it is located within a densely populated area and can easily be visited. SOFIA offers a unique opportunity for the public to become involved in front-line astronomy and to see how a modern astronomical observatory actually works. The aircraft even provides enough space to welcome a few guests on board during each flight. The DLR has decided to support the EPO program with independent but related activities in Germany. Presenting SOFIA at exhibitions, conferences, and in the media, preparing printed information and web pages (SOFIA 1999), are part of this PR program.

Figure 11.
Predicted SOFIA image quality. The spatial resolution SOFIA can achieve is roughly represented by the *50% enclosed light diameter* curve.

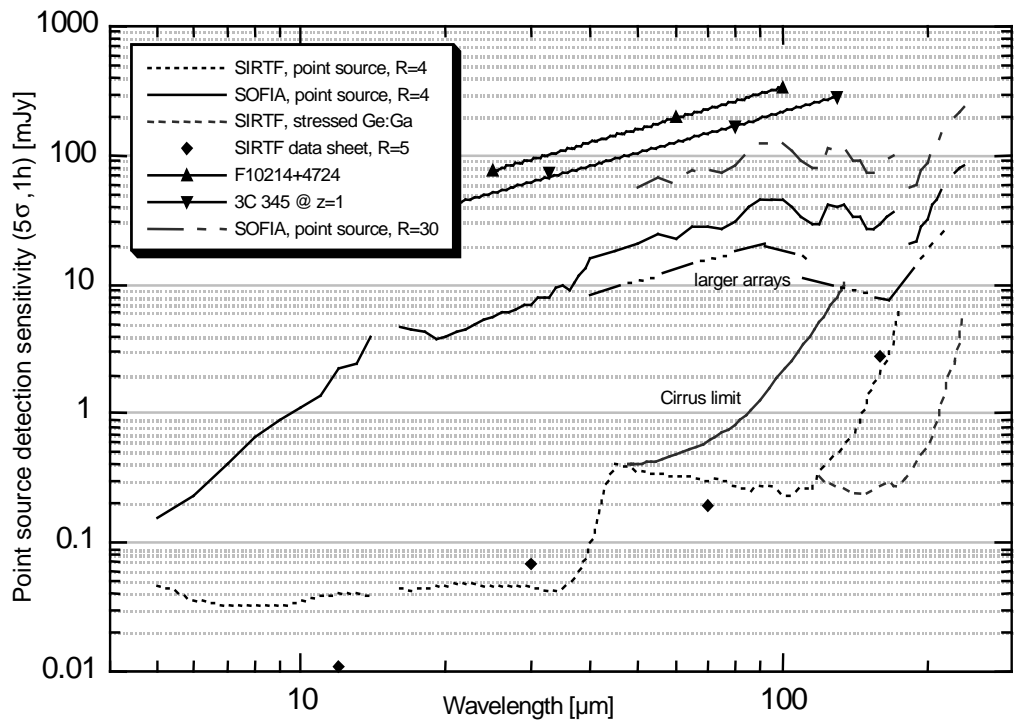

Figure 12. SOFIA's point source sensitivity for broad band imaging and low-resolution spectroscopy. Curves for the SIRTF mission are given for comparison. The triple-dot-dashed curve denotes the virtual increase of sensitivity of SOFIA compared with SIRTF

that can be achieved by using a 64 x 64 pixel Ge:Ga and a 32 x 32 stressed Ge:Ga array. The "Cirrus limit" curve is a rough estimate (see e.g. Herbstmeier et al. 1998 for more details).

## 6. Expected Performance

### 6.1 Spatial resolution

The spatial resolution achievable with SOFIA is displayed in Figure 11. The curves include the effects of wind turbulences and diffraction. The 50% enclosed light diameter curve represents about the limit of the spatial resolution achievable with SOFIA's 2.5m aperture. SOFIA will be diffraction limited from about 10 µm on. Below 5µm, the seeing is dominated by wind turbulence within and in the outer vicinity of the cavity. Between 30µm and 300µm SOFIA's spatial resolution will be unprecedented and the highest achieved so far. It will be more than 4 and 3 times higher than that of ISO and SIRTF, respectively.

### 6.2 Sensitivity

The results reported here are based on sensitivity models for SOFIA covering a range of observing scenarios. A detailed description of these models is given by Krabbe et al. (1998).

6.2.1 Broad band imaging

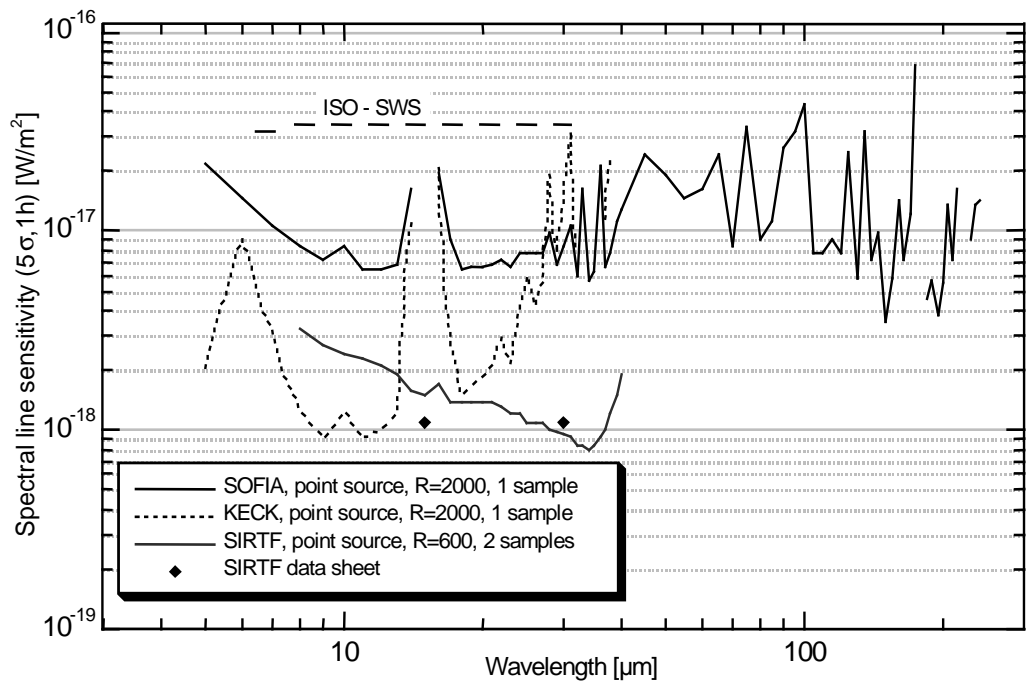

Figure 13. Calculated SOFIA sensitivity for R ~ 2000 spectroscopy of point-like sources as compared with the Keck telescope and the projected SIRTF satellite. The dashed curve shows the practical detection limit of the ISO short wavelength spectrometer SWS. The features in the curves give an impression of the expected variations of the spectral sensitivity due to the atmospheric transmission profile.

For the point source sensitivity an image scale of 2 pixels per $\lambda/D$ and 8 pixels per integrating aperture was assumed. SOFIA's $5\sigma$ 1h photometric sensitivity ranges from about 200µJy at 5µm to about 40 mJy at 100µm (Figure 12). The expected sensitivity of SIRTF is given for comparison. Beyond 100µm galactic cirrus will degrade SIRTF's point source sensitivity (e.g. Herbstmeier et al. 1998, Kawara et al. 1998) to the extend that it might come close to a factor of 2 of the SOFIA limit. Given the potential of flying larger arrays on SOFIA in the future, SOFIA's effective sensitivity to faint sources compares well with SIRTF beyond 100µm.

SOFIA's photometric sensitivity on extended sources was calculated with the same pixel scale as above. For R=5, SOFIA's $5\sigma$, 1h, 1pixel sensitivity ranges from below 100MJy/sr at MIR to about 25MJy/sr beyond 100µm.

6.2.2 Spectroscopy

SOFIA's predicted sensitivity for medium resolution spectroscopy on point like targets is plotted in Figure 13. The spectrum is Nyquist sampled, whereas the pixel size in the spatial direction is λ/D. The features in the curve are dominated by the atmospheric transmission profile and give an impression of the expected variations of the spectral sensitivity in the various parts of the spectrum. The derived 5σ, 1h averaged sensitivities for SOFIA all lie in the range $1 - 2 \times 10^{-17}$ W/m$^2$, about 3 times better than ISO-SWS (see A.&A. Lett., vol. 315, (1996), special issue).

## 7. Scientific Potential

With SOFIA's unique scientific capabilities we expect significant contributions to many areas of modern astronomical research. Four topics will be addressed here: The evolution of galaxies, the formation and evolution of stars and the role of the interstellar medium (ISM), and our planetary system.

### 7.1. Evolution of galaxies

With the advent of larger ground-based and space borne telescopes, covering a broad spectral range, galactic evolution has grown into a major research area. SOFIA will significantly contribute to, or solve, several of the key issues, some of which are briefly discussed here.

7.1.1 Interacting Galaxies & Clusters

Evidence is growing that interaction between galaxies may be the dominant driving force behind the evolution and diversification of galaxies (e.g. Moore et al. 1996). Most of the luminous and ultraluminous galaxies including quasars and other AGN's known today are part of interacting or merging binaries or multiple galactic systems (e.g. Sanders et al. 1988, Disney et al. 1995, Miles et al. 1996, Sanders & Mirabel 1996 and references therein). Such interaction almost always destroys the previous structure of the interstellar gas and dust in the participating galaxies. Angular momentum transfer easily triggers the infall of giant molecular gas and dust clouds towards the central regions of the galaxies where they collide with each other and eventually start a powerful starburst (SB). New populations of stars are generated by such events and the morphological type of the galaxy may change completely. Numerical calculations show that the majority of cluster galaxies have already gone through one or more such interactions (Moore et al 1996).

Although interaction processes between galaxies leading to SB events seem to be crucial for the understanding of galactic evolution as a whole, our knowledge of the phenomena involved is yet preliminary. The reason is that most signposts for such interaction processes show up first and most clearly in the infrared. Existing stars are not very much affected by interactions since stellar collisions are negligible. Collisions between molecular clouds, however, generate a plethora of shocks, turbulences, heating of dust particles and excitation, most of which can only be traced in the infrared. The possible aftermath of such an event, a powerful SB or eventually an active galactic nucleus (AGN), will create a lighthouse beam in the far infrared in the form of the huge spectral bump in the wavelength range 60µm to 100µm. It resembles reradiation of dust exposed to the powerful source of energy in the nucleus and it's bolometric luminosity can exceed that of our galaxy by more than 4 orders of magnitude. Since the detection of such galaxies by IRAS (Soifer, Houk, & Neugebauer 1987 and references therein), SOFIA will become the first MIR/FIR observatory sensitive enough to efficiently observe such targets. It will thus be possible for the first time:

- to analyze these events on a more statistical basis in particular in isolated clusters,
- to determine the spectral energy distribution (SED) of the sources involved,
- to measure their dust temperatures and separate components of different temperatures,
- to spatially resolve and investigate nearby merging systems in the MIR & FIR,
- to detect interacting systems at larger redshifts.

7.1.2. Medium and high-z galaxies

If all the galaxies we observe today underwent violent episodes to form their different populations of stars, we should be able to see the footprints of those events in the history of the observable universe. Looking at the spectral energy distribution (SED) of normal and luminous infrared galaxies (LIRG) it is obvious that galaxies with high redshifts and the highest luminosities will be detected most easily in the MIR and FIR spectral range. There has already been a successful attempt using the ISO satellite at 6µm and 15µm to detect and identify luminous infrared galaxies in the Hubble Deep Field (Rowan-Robinson et al. 1997). Nine of the 10 objects with predicted 100µm fluxes are either stronger than SOFIA's (5σ, 1h, 40 mJy) sensitivity or fainter within a factor of 3, which still is practicable.

With SOFIA's predicted sensitivity one can determine the largest distance at which galaxies of a given luminosity can be detected with SOFIA at 100µm. Figure 14 shows the result for 8 galaxies including starbursts, Seyferts, interacting galaxies, and quasars. Their apparent luminosity range is $10^{10}$-$2 \cdot 10^{14}$ $L_O$ and their redshifts go out to z = 2.3. The straight line indicates SOFIA's (5σ, 1h) detection sensitivity: A galaxy will be detected if its luminosity $L > 2.3 \cdot 10^4 L_O (\frac{D_L}{Mpc})^{2.28}$, where $D_L$ is the luminosity distance in Mpc.

Ultraluminous infrared galaxies (ULIRG) with $L = 10^{12} L_O$ can be observed out to z ~ 1 and galaxies comparable to the most luminous galaxies known can be detected out to z ~ 3. The sensitivity is high enough to study the evolution of IR bright galaxies in clusters at around z = 0.3 ~0.4, which are typical distances for so called "Butcher-Oemler" clusters (Couch et al. 1994). Although the atmospheric background will limit SOFIA's sensitivity generally to galaxies of z ≤ 1, the spatial resolution of SOFIA between 30µm and 250µm will be unprecedented. It will allow us

- to disentangle crowded field much better than ISO and SIRTF,
- to detect and complete the sample of IR bright galaxies out to a higher redshift, at wavelengths where ISO and SIRTF are confusion limited,
- to identify IR counterparts of deep optical and NIR images of the Hubble Space Telescope (HST),
- to search for galaxies in an early phase of their evolution (e.g. blue galaxy clusters),
- to determine their SEDs if their 100µm flux density exceeds ~100mJy,
- to compare the interaction rate and characteristics of the star formation events on different z scales.

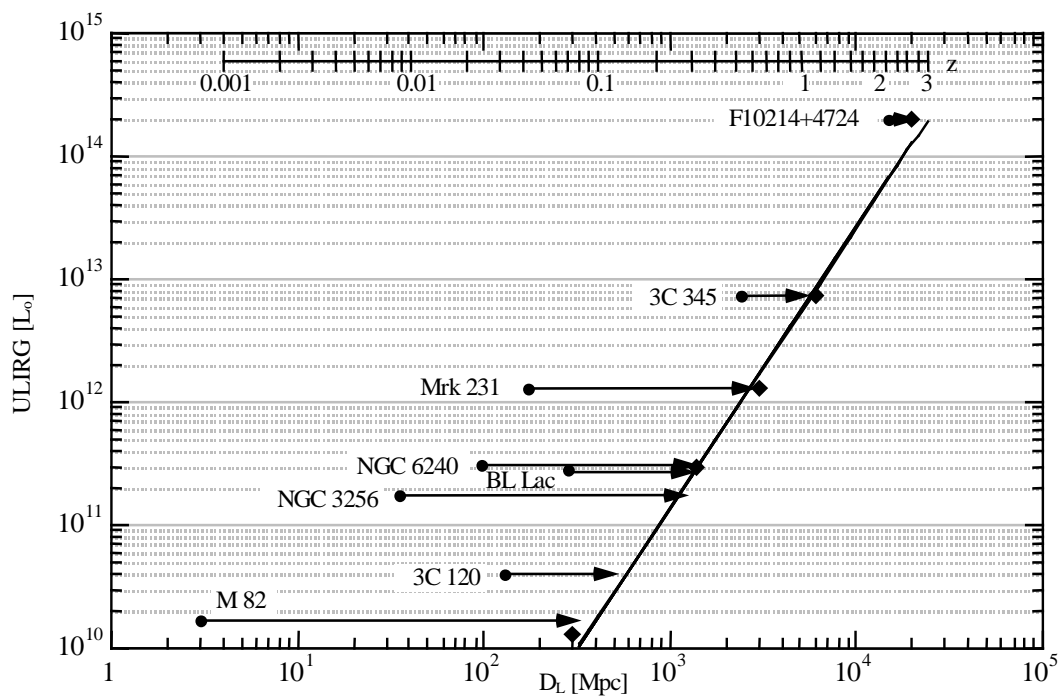

Figure 14. The line indicate SOFIA's (5σ, 1h) faint galaxies detection limit constrained by their distances and their infrared luminosities. The arrows indicate how far galaxies of known distance can be redshifted before they disappear for SOFIA at 100µm.

In addition it is very likely to observe or even discover new classes of objects. The so-called ERO's, for example, seem to be a population of very red galaxies discovered in the near infrared (e.g. Elston, Rieke, & Rieke 1988, 1989; Cimatti et al. 1997). Very little is yet known about their nature, their SED's, their sources of far infrared emission, and even their redshifts. Graham & Dey (1996) speculate that their 200µm flux density may reach 100mJy, well within the range observable with SOFIA.

The nature of the recently discovered extragalactic FIR-background (e.g. Guiderdoni et al. 1997, Harwit 1999) is still under debate. SOFIA's cameras will be ideally suited for follow-up observations and should be able to prove whether the FIR-background is being emitted by high-z galaxies and what the characteristics of those objects are.

### 7.1.3 ISO follow-up

The Infrared Space Observatory has already delivered a wealth of data, in particular many new results on the evolution and nature of starburst galaxies, ULIRG's and active galactic nuclei (AGN)(see A.&A. Lett., 315, (1996), special ISO issue). Providing a mirror diameter more than 4 times larger than ISO, SOFIA is ideally suited to follow-up on those targets in much greater spatial detail and with an improved spectroscopic sensitivity (see section 6).

In the case of SB galaxies it will thus be possible to proceed from ISO's determination of integral SB parameters to a spatially detailed analysis in the waveband between 15µm and 40µm on many sources. Individual SB regions will be localized and their dust temperature and distribution will be determined. Ratios of spectral line maps, such as [NeIII]/[NeII] (15µm/13µm) or [SIV]/[SIII] (18µm/33µm), will be used to determine the radiation field, the excitation conditions and the dynamics in these regions and thus analyze the SB parameters. Dust components of low temperature will be searched for. These trace cold masses, which are important contributors to the galaxies total mass budget; the latter is still not well defined for the majority of the galaxies.
The Seyfert activity still bears two unsolved questions: The existence of the so called *unified scheme* that would allow us to understand different types of Seyfert activity within the frame of a single model, and the type of physical and/or evolutionary connection between Seyfert- and SB activity which are often closely associated with each other. The study of the small Seyfert nuclei, however, is difficult. AGN are often so closely associated with SBs or even surrounded and obscured by them that the nuclei themselves are not directly accessible in the optical and near-infrared (NIR) or their emission cannot be separated from the SB component. In the MIR, however, the situation is much improved because the combined emission of starlight and dust emission from the SBs reaches a minimum there, whereas the emission of the active nuclei has its maximum.

SOFIA's excellent spatial resolution will enable us for the first time to isolate the nuclear emission reliably from the rest of the galaxy, to study the nuclear energetic properties, the SED of the hot dust and its heating mechanism as a function of different Seyfert types and SB environments. High excitation lines like [NeV] (24µm) and [OIV] (26µm) are much less obscured compared to the NIR and trace the narrow line regions.

### 7.2 Star formation and evolution

Our understanding of the processes involved in the formation of low- and high-mass stars has made great progress within the last years. Stars generally form through a fragmented collapse of a molecular cloud as a group or cluster. This simple picture seems to be generally true but bears many unsolved details. The complete initial mass function of such groups is still unknown. The formation of low-mass stars is presently unclear. What are the lower and upper mass cut-offs? What role does the environment of the collapsing clouds play, for example, its radiation field and winds? How does the formation of the high-mass stars in a group influence the ongoing star formation. What is the role of metallicity in the star formation process?

Unfortunately, the early stages of star formation, where all these processes run and stellar masses are determined, cannot be observed easily. The reason is that the spectral emission of warm and hot gas clouds with temperatures between 20K and 200K peaks at wavelengths between 150µm and 15µm, mostly inaccessible from the ground. Airborne or space borne experiments, on the other hand, are or have been limited in their spatial resolution to several 10arcsec and are thus not able to spatially resolve star forming regions well enough. In addition, many of these star forming regions are heavily obscured by dense foreground dust clouds with $A_v > 100$. Such foreground extinctions make observation difficult even in the near infrared (e.g. Orion KL region). The MIR & FIR instruments foreseen for SOFIA will be sensitive enough and provide a high enough spatial resolution (about 5arcsec at 60µm) to solve these issues.

SOFIA will allow us to spatially separate the protostars and study them individually. Mass functions of different regions, diverse in geometry and metallicity, will be compared. Deeply embedded high-mass stars will be searched for and their environment studied. Cameras onboard SOFIA will be able to map out the environment of pre-main-sequence objects like TTauri, HH-objects, and Ae/Be stars to study their expanding shells, envelopes or outflows which are probably remnants from the parent molecular clouds. The spectral energy distribution of such shells will be determined with low-resolution spectroscopy. These studies are crucial for our understanding of how stars and planetary systems form, and how our solar system may have formed.

Circumstellar concentrations of dust have been discovered around several evolved stars, β Pic being the most well known example. It may well be possible that circumstellar dusty disks are fairly common since they have also been discovered around such ordinary stars as α Lyrae. The study of such disks or concentrations, their shape and statistics, is essential for the understanding of the origin and formation of planetary systems which is linked to the question of our own existence.

### 7.3 Interstellar Matter

The ISM reflects the global history of a galaxy (e.g. interaction) as well as local phenomena (e.g. star formation, radiation field). As such it links the global galactic evolution with the localized stellar evolution. An understanding of the formation and evolution of galaxies as well as the formation of stars provides insight into the physics and chemistry of the ISM. The complex structure of the ISM is a consequence of the fact that the ISM is an open system that tries to balance itself according to the physical and chemical conditions that prevail locally. Since most of the ISM is cold or, in the presence of a radiation field, warm, most of its continuum radiation and spectral line emission emerges in the MIR, FIR, and Submm. SOFIA is thus the only observatory ideally suited to study these processes.

SOFIA will allow us, to understand the chemical fractionation of matter within the galaxy. Photon dominated regions (PDRs) are the stages for the dynamical interaction of star formation and its radiation field with the ISM. The molecular clouds absorb part of the radiation field, and produce a diversity of species that only exist in a well-defined local equilibrium. The equilibrium mainly depends on the distance from the radiation source (mostly a hot star) and its radiated spectrum, and on the density, temperature, chemical composition, and geometry of the molecular cloud. The diffuse interstellar radiation field plays the dominant role in the chemistry of more isolated MCs. High resolution heterodyne spectroscopy will be used to obtain spectral maps of molecules (e.g. HD, $H_2O$, CO, HCN), atoms (e.g. H, C, O), ions ($H^+$, $C^+$, $Ne^+$), and radicals (e.g. $HCO^+$, $CO^+$, $OH^+$) in molecular cloud regions.

Massive stars dominate their environment by their strong and energetic radiation field. At later stages of their evolution, stellar matter is fed back into the interstellar region by mass loss of high mass stars (e.g. LBVs), stellar winds of evolved AGB stars, planetary nebulae, stellar jets, and violent supernova (SN) events. These stars eject processed, i.e. metal enriched, material back into space. The interaction of this material with the surrounding interstellar matter (ISM) leads to shock fronts which heat and destroy the dust and excite characteristic spectral lines. The study of the composition of the reprocessed material and its interaction with the ISM will strongly impact on our understanding of the chemistry and dynamics of the ISM as well as on the galactic chemical evolution.

### 7.4 Solar System

Pre-solar nebulae, circumstellar dust shells and dust rings are the building blocks of solar systems and it is believed, are closely related to the origin of life. Our solar system is the obvious place to study the conditions and processes which lead to the formation of planets and the evolution of their surfaces and atmospheres, and to search for traces of pre-biological chemistry. It is therefore essential to study primordial bodies like comets and asteroids as well as the evolved planets and their satellites.

7.4.1 Comets and asteroids

Comets are composites of ice and dust with diameters up to several 10s of km, which orbit in the Kuiper Belt and Oort cloud. If they are somehow deflected towards the inner solar system, their surfaces are exposed to intense solar radiation and the ices start to vaporize. Since this interaction with solar radiation is very probably the first chemical processing in billions of years, comets provides us with information about the composition of matter at the time of their formation. Their different compositions seem to represent different regions of the pre-solar nebula. Some are rich in silicates, others in reduced carbon or hydrocarbons; others show various proportions of minerals like olivine. Most of them seem to be rich in water. Spectroscopic information of the most important volatile components of comets like $H_2O$ and $CO_2$ or organic C-X stretch bands cannot be obtained from the ground since the absorption from water and carbon dioxide in the lower atmosphere blends out those bands. SOFIA is therefore ideally suited to address those questions.

Most asteroids orbit in the main belt between the orbits of Mars and Jupiter. Asteroids are thought to be remnant material from the processes of formation and initial development of planets and therefore an important source of information on conditions in the early solar system. Apart from the scientific question of their origin and composition, the so-called near-Earth asteroids (NEAs) are also interesting as potential sources of raw materials for future generations and as targets for current space missions such as NEAR and DS1. Observations of asteroids in the MIR and FIR with SOFIA will provide crucial information on their physical characteristics, such as sizes, albedos and the thermal properties of their surfaces (see, e.g. Harris et al.1998). Well-studied asteroids with accurately known characteristics will serve as important photometric standards for SOFIA in the MIR and FIR, as in the case of ISO (Mueller and Lagerros 1998)

7.4.2 Planets and their satellites

The composition of the planets is directly linked to the distribution of matter and volatiles during the formation of our solar system. The radial partitioning of matter during the formation of planets should reflect itself in the different compositions of the planets. For the big outer planets, which are difficult to investigate, models range between two extremes: outer planets and their satellites consist of (1) water, methane, and ammonia or (2) carbon monoxide, water, and molecular nitrogen. In the currently accepted model, the giant planets consist of (2), while their satellite's compositions are dominated by (1). The satellites of the outer planets may retain surface spectral signatures of the primordial partitioning of these constituents. It is therefore important to determine the composition of the volatile-rich outer planet satellites and Pluto, the composition of their atmospheres (if any) and the nature of the surface-atmosphere interactions.

The spatial-seeing-limited NIR resolution of SOFIA is about 2 arcsec (Figure 11), high enough to spatially resolve several of the planetary disks (Table 4) and study their zonal atmospheres.

Table 4   SOFIA's spatial NIR resolution on planets and satellites

| Body | Diameter [km] | 2" resolution on body [km] | No. of resolution elements across diameter |
|---|---|---|---|
| Mars | 6800 | 760 | 9.0 |
| Jupiter | 1428000 | 6100 | 23.4 |
| Saturn | 120000 | 12400 | 9.7 |
| Uranus | 52000 | 26400 | 2.0 |

On Mars, SOFIA will be able to study the transport of the volatiles $CO_2$ and perhaps $H_2O$ between the polar caps and the equator region due to seasonal cycles. On Jupiter, SOFIA can spectroscopically map the cloud band system and its variations in NIR and MIR lines. SOFIA will be able to resolve the Great Red Spot, which color is still unexplained.
Stellar occultations of planets and satellites have been a very valuable source of information on planetary atmospheres and rings during the KAO flights. SOFIA's mobility is ideally suited for observations of these occultations, which provide a spatial resolution of only a few kilometers. In 1997, the rings of Uranus were discovered with the KAO during a stellar occultation. SOFIA not only will provide a much-improved sensitivity but also long term monitoring of changes within these atmospheres and rings.

Table 5  Approved first light instruments for SOFIA

| Name of Instrument | PI | Institute | Type of Instrument | Facility, PI, or Special Class |
|---|---|---|---|---|
| HOPI | E. Dunham | Lowell Observatory | Occultation CCD Photometer / Imager | Special Class |
| AIRES | E. Erickson | NASA – ARC | Echelle Spectrometer 17 to 210 microns R=10,000 | Facility Instr. |
| HAWC | D.A. Harper | Univ. of Chicago | Far Infrared Bolometer Camera 30-300 microns | Facility Instr. |
| FORCAST | T. Herter | Cornell | Mid IR Camera 5-40 microns | Facility Instr. |
| EXES | J. Lacy | Univ. of Texas | Echelon Spectrometer 5-28 microns R=1500 and R=100,000 | PI Instr. |
| FLITECAM | I. McLean | Univ. of California, Los Angeles | Near IR Test Camera 1-5 microns | Facility Instr. |
| CASIMIR | J. Zmuidzinas | Caltech | Heterodyne Spectrometer 250-600 microns | PI Instr. |
| SAFIRE | S. Moseley | NASA-GSFC | Imaging Fabry-Perot 145 – 655 microns | PI Instr. |
| GREAT | R. Güsten | MPIfR, Bonn; Uni Köln; DLR, Berlin | Heterodyne Spectrometer 75 - 250 microns | PI Instr. |
| FIFILS | A. Poglitsch | MPE, Garching Uni Jena | Field Imaging Far IR Line Spectrometer 40-350 microns | PI Instr. |

## 8. First light instruments

After a phase of thorough testing and adjusting of the observatory, the regular guest-observing program will probably start at the end of 2002. Ten scientific instruments have already been selected and approved and should be available at first light (Table 5). Four of them are American facility instruments, one fulfills special purposes and 4-5 are principal investigator (PI) instruments. Two PI instruments are provided by German institutions. FLITECAM, a facility instrument, is also the official test camera used to check all optically related properties of the telescope, including pointing and tracking, chopping and nodding, wind loads and other sources of vibrations, the optical quality and the emissivity, telescope temperature distribution and others. A more detailed description of the instruments is given in the web pages (SOFIA 1999). Three additional German astronomical instruments have officially been proposed (funds pending): SPICA (DLR Berlin, a spectral-photometric IR camera), STAR (Univ. Köln, a Heterodyne receiver in the Terahertz band), and IRHEAT (Univ. Köln).

*Acknowledgements*
The German SOFIA project is funded by German Aerospace Center (former DARA) and managed by A. Himmes and W. Klinkmann. We thank J. Wolf, A. Harris, and Jacqueline Davidson for critical reading and for helpful comments.